\documentclass[12pt,preprint]{aastex}
\usepackage{natbib}
\usepackage{rotating}
\usepackage{mathptmx}

\newcommand{\kms}{kms$^{-1}$}

\newcommand{\lta}{\raisebox{-0.6ex}{$\,\stackrel
{\raisebox{-.2ex}{$\textstyle <$}}{\sim}\,$}}

\shorttitle{Cosmological constraints on the proton-to-electron mass ratio}
\shortauthors{S. P. Ellingsen et al.}

\begin{document}

\title{First cosmological constraints on the proton-to-electron mass ratio from observations of rotational transitions of methanol}

\author{S. P. Ellingsen}
\affil{School of Mathematics and Physics, University of  Tasmania, 
  Private Bag 37, Hobart, TAS 7001, Australia}
\email{Simon.Ellingsen@utas.edu.au}
\author{M. A. Voronkov, S.L. Breen}
\affil{CSIRO Astronomy and Space Science, Australia Telescope National Facility, PO Box 76, Epping, NSW 1710, Australia}
\author{J.E.J. Lovell}
\affil{School of Mathematics and Physics, University of  Tasmania, 
  Private Bag 37, Hobart, TAS 7001, Australia}
  
\begin{abstract}
We have used the Australia Telescope Compact Array to measure the absorption from the $2_{0} \rightarrow 3_{-1}E$ 12.2~GHz transition of methanol towards the $z$=0.89 lensing galaxy in the PKS$\,$B1830$-$211 gravitational lens system.  Comparison of the velocity of the main absorption feature with the published absorption spectrum from the $1_{0} \rightarrow 2_{-1}E$ transition of methanol shows that they differ by $-0.6 \pm 1.6$ \kms .  We can use these observations to constrain the changes in the proton-to-electron mass ratio $\mu$ from $z$=0.89 to the present to $0.8 \pm 2.1 \times 10^{-7}$.  This result is consistent, and of similar precision to recent observations at $z$ = 0.68 achieved through comparison of a variety of rotational and inversion transitions, and approximately a factor of 2 better than previous constraints obtained in this source.  Future more sensitive observations which incorporate additional rotational methanol transitions offer the prospect of improving current results by a factor of 5-10.
\end{abstract}


\section{Introduction}

Astrophysical observations provide one of the most sensitive methods for searching for possible temporal or spatial changes in the fundamental constants \citep{Uzan11}.  Under the standard model of particle physics these constants are not expected to change, however, the detection of variations would be consistent with some anthropic models, as well as indicating the need for new physics beyond general relativity and the standard model.  A large number of observations have been undertaken to attempt to measure, or constrain changes in the fine-structure constant $\alpha=e^2/\hbar c$ \citep[e.g.][]{Webb+99,Levshakov+06} and the proton-to-electron mass ratio $\mu=m_p/m_e$ \citep[e.g.][]{King+08,Henkel+09,Kanekar11,Muller+11}, or combinations of $\alpha$, $\mu$ and the nuclear $g$-factor $g_n$ \citep[e.g.][]{Kanekar+05,Kanekar+10}.  Although some studies have claimed marginal detections of changes in either $\alpha$, $\mu$ or the combinations, none have yet been confirmed through additional independent observation.  Changes in the constants are measured through comparison of the frequency ($\nu$) of two transitions which have different sensitivities to variations in one or more of $\alpha$, $\mu$, or $g_n$.
\begin{equation}
\frac{\Delta \nu}{\nu} = K_{\alpha} \frac{\Delta \alpha}{\alpha} + K_{\mu} \frac{\Delta \mu}{\mu} + K_{g} \frac{\Delta g_n}{g_n}, 
\end{equation}
where the coefficients $K_{\alpha}$, $K_{\mu}$ and $K_g$ represent the sensitivity of the particular transition to changes in that constant.

The proton-to-electron mass ratio will change if there are differences in the spatial or temporal scale of variations in the strong nuclear force compared to the electromagnetic force, and recent observational studies have focused on using molecular observations at radio wavelengths to search for changes in $\mu$.  Rotational transitions of molecules are generally more sensitive to changes in $\mu$ than ro-vibrational transitions of H$_2$, and recent studies have focused on comparison of absorption in NH$_3$ inversion transitions with rotational transitions from molecules such as CS and HCO$^+$ \citep{Henkel+09,Kanekar11}.  This is because ammonia inversion transitions are more sensitive to changes in $\mu$ than the rotational transitions of most molecules, with $K_{\mu}$ = -4.46 \citep{Flambaum+07}.  However, recent investigations have revealed that the different rotational transitions of the methanol molecule have a larger and more varied sensitivity to changes in $\mu$ than any other molecule identified to date \citep{Jansen+11,Levshakov+11}, with some transitions having approximately an order of magnitude greater sensitivity than the ammonia inversion transitions.  Furthermore, in the local universe methanol emission is commonly observed from high-mass star formation regions in the form of masers, and thermal emission from hot cores, and absorption is detected towards cold clouds in the foreground of continuum sources \cite[e.g.][]{Menten91,vanderTak+00,Peng+92}.  Variations in $\mu$ with density are a prediction of chameleon-like scalar fields, which are one of the mechanisms which can produce dark energy, and the densities in interstellar molecular clouds are many orders of magnitude lower than can be achieved in the laboratory.  Observations of methanol masers within the Milky Way have recently been used to constrain spatial variations in the proton-to-electron mass ratio to $\Delta \mu/\mu < 8.1 \times 10^{-8}$ (3-$\sigma$) \citep{Levshakov+11,Ellingsen+11}.

Until very recently methanol emission and absorption had only been detected in nearby galaxies, however, a spectral scan from 30--50~GHz in the molecular absorption system towards the gravitationally lensed quasar PKS\,B1830$-$211 yielded the first detection of methanol at cosmological distances \citep{Muller+11}.  These observations detected methanol absorption from a single transition, the $1_{0} \rightarrow 2_{-1}E$ which has a rest frequency of approximately 60.5~GHz, which for a redshift of $z$=0.89 is observed at a frequency of approximately 32.1~GHz.  For sources in the local Universe the emission/absorption from this transition is within a region of the electromagnetic spectrum which cannot be studied from the Earth's surface due to absorption by atmospheric O$_2$.  Because of this there are no previous studies of this transition.

PKS\,B1830$-$211 is a very well studied gravitational lens system, with a quasar at a redshift of $z$=2.507 \citep{Lidman+99} and the primary lensing galaxy at a redshift of 0.88582 \citep{Wiklind+96}.  There is also evidence for a second galaxy along the line of sight at a redshift of 0.19 \citep{Lovell+96}.  The two main components of the gravitationally lensed quasar are separated by nearly 1$\arcsec$ on the sky, and at frequencies \lta 10~GHz part of the quasar jet forms an Einstein ring \citep{Jauncey+91}.  The quasar has a steep spectrum jet and an optically thick core component.  The south-western core component has an angular size which varies as wavelength squared, consistent with interstellar scattering in the lensing galaxy \citep{Jones+96}.  The majority of the radio continuum emission is contained in the two compact components which lie to the north-east and south-west of the centre of the Einstein ring, with the north-eastern component being the stronger of the two.  The lensing galaxy at $z$=0.89 is an Sb or Sc spiral seen nearly face-on \citep{Winn+02}, with the line-of-sight of the south-western component intersecting the galaxy at a galacto-centric radius of approximately 2~kpc, and the north-eastern at a radius of $\sim$4~kpc.  The strongest molecular absorption is seen towards the south-western component, while the strongest H{\sc i} absorption is seen towards the north-eastern component \citep{Carilli+98,Chengalur+99}.  The velocity offset between the absorption from the south-west and north-east components is approximately 150~\kms.  The molecular absorption in PKS\,B1830$-$211 has been the subject of numerous studies with more than 30 different molecular species detected and multiple complexes spanning a velocity range of nearly 500~\kms\/  \citep[see][and references therein]{Muller+11}. To date, no molecular absorption has been detected towards the second line-of-sight galaxy at $z$=0.19.

The 12.2 GHz ($2_{0} \rightarrow 3_{-1}E$) transition of methanol lies in the same series as the 60.5 GHz and this transition is well studied as it shows both strong maser emission in Galactic high-mass star formation regions \citep[e.g.][]{Breen+11} and absorption in cold clouds \citep{Peng+92}.  \citet{Ellingsen+11} suggested that absorption from the 12.2~GHz transition was likely to be detectable in this source and noted that the redshifted frequency of approximately 6.45~GHz lay within the frequency range of both the Australia Telescope Compact Array (ATCA) and the EVLA.  Observations of this transition, combined with the published spectra for the 60.5 GHz transition present a unique opportunity to utilise the high sensitivity of rotational transitions of methanol to search for possible variations in $\mu$ at cosmological distances.  Here we present the first observations of the $2_{0} \rightarrow 3_{-1}E$~12.2 GHz transition of methanol towards the $z$=0.89 molecular absorption system in PKS\,B1830$-$211.

\section{Observations} \label{sec:obs}

The observations were made using the ATCA in Director's time allocations on 2011 November 4, November 18 and December 7 (project code CX223).  The Compact Array Broadband Backend (CABB \citet{Wilson+11}) was configured with a 2 $\times$ 2 GHz bands covering the frequency ranges 4.876-6.924 GHz and 8.667-10.715 GHz.  Spectral zoom bands were centred on frequencies of  6.547 GHz and 10.587 GHz, corresponding to the approximate observing frequencies of the $2_{0} \rightarrow 3_{-1}E$ (12.2 GHz) and $2_{1} \rightarrow 3_{0}E$ (19.9 GHz) transitions of methanol at a redshift $z$ = 0.89.  The rest frequencies assumed throughout this work were 12.178597 and 19.9673961 GHz \citep{Muller+04}.  For the November 4 observations 16 $\times$ 1 MHz spectral zooms were concatenated with overlaps (for each of the two transitions ), to produce spectra with 17408 spectral channels across a bandwidth of 8.5 MHz, corresponding to a spectral resolution of 488 Hz.  For the November 18 and December 7 observations a single 64 MHz spectral zoom with 2048 channels was used for each transition, corresponding to a spectral resolution of 31.25 kHz (i.e. a factor of 64 coarser than the November 4 observations).  The details of the observations are summarised in Table~\ref{tab:obs}.

The primary target for the observations was the gravitationally lensed quasar PKS\,B1830$-$211, which has previously been observed to exhibit molecular absorption from a wide range of molecules from the lensing galaxy at $z$ = 0.89.  The pointing centre for the observations was $\alpha = 18^{\mbox{h}}33^{\mbox{m}}39.90^{\mbox{s}}$ ; $\delta = -21\arcdeg03\arcmin40.0\arcsec$ (J2000).  The observations were constructed as a series of 15 minute scans on PKS\,B1830$-$211, interleaved with 2 minute observations of PKS\,B1908$-$201 which were used for phase calibration.  For each session observations of PKS\,B1921$-$293 were undertaken for bandpass calibration and of PKS\,B1934$-$638 for primary flux density calibration.  The data were reduced using the {\sc miriad} software package \citep{Sault+95}, applying the standard techniques for spectral line observations, except for the bandpass calibration for the second and third observing sessions.  For the observations with a 64 MHz spectral zoom the bandpass calibration was achieved using continuum data for PKS\,B1830$-$211extracted using the task {\em uvlin}.  This effectively fits a polynomial to the PKS\,B1830$-$211 continuum and resulted in a flatter and lower-noise bandpass solution than using PKS\,B1921$-$293.

For each of the three observing sessions we extracted a spectrum from the {\em uv} data by vector averaging at the position of the south-west component of the PKS\,B1830$-$211 gravitational lens ($\alpha = 18^{\mbox{h}}33^{\mbox{m}}39.866^{\mbox{s}}$ ; $\delta = -21\arcdeg03\arcmin40.45\arcsec$ (J2000) \citet{Subrahmanyan+90}).  This is an offset on the sky of -0.196$\arcsec$ in right ascension and -0.450$\arcsec$ in declination from the pointing center of the observations.   To enable the data collected on different days to be averaged, and for comparison with molecular absorption observed in other transitions the observed sky frequency was corrected for the Doppler shift of PKS\,B1830$-$211 in the Barycentric frame (on a per-visibility basis within {\sc miriad}).  The velocity of the array towards PKS\,B1830$-$211 in the Barycentric frame is listed in Table~\ref{tab:obs} for each observing session.  The spectra for each session were produced with the same spectral resolution (62.5 ~kHz) and aligned in frequency.  An average spectrum was constructed from the weighted mean of the three days, with the weighting proportional to the time on-source for each day.  Prior to averaging, the flux density scale of the spectra were normalised to the continuum level observed at the position of the south-west component.  The continuum level varied between sessions (see Table~\ref{tab:obs}), both due to the differing angular resolution of the observations, and due to to the strong intrinsic variability well known in this source \citep{Lovell+98}.  We then transformed the average spectrum to the velocity relative to a redshift of $z$=0.88582, the redshift of the absorption towards the south-western component of the lens \citep{Muller+06}.  The velocity resolution of the 12.2 GHz spectrum in the rest frame of the absorbing system is 2.9~\kms .  

\section{Results} \label{sec:results}

\begin{figure}
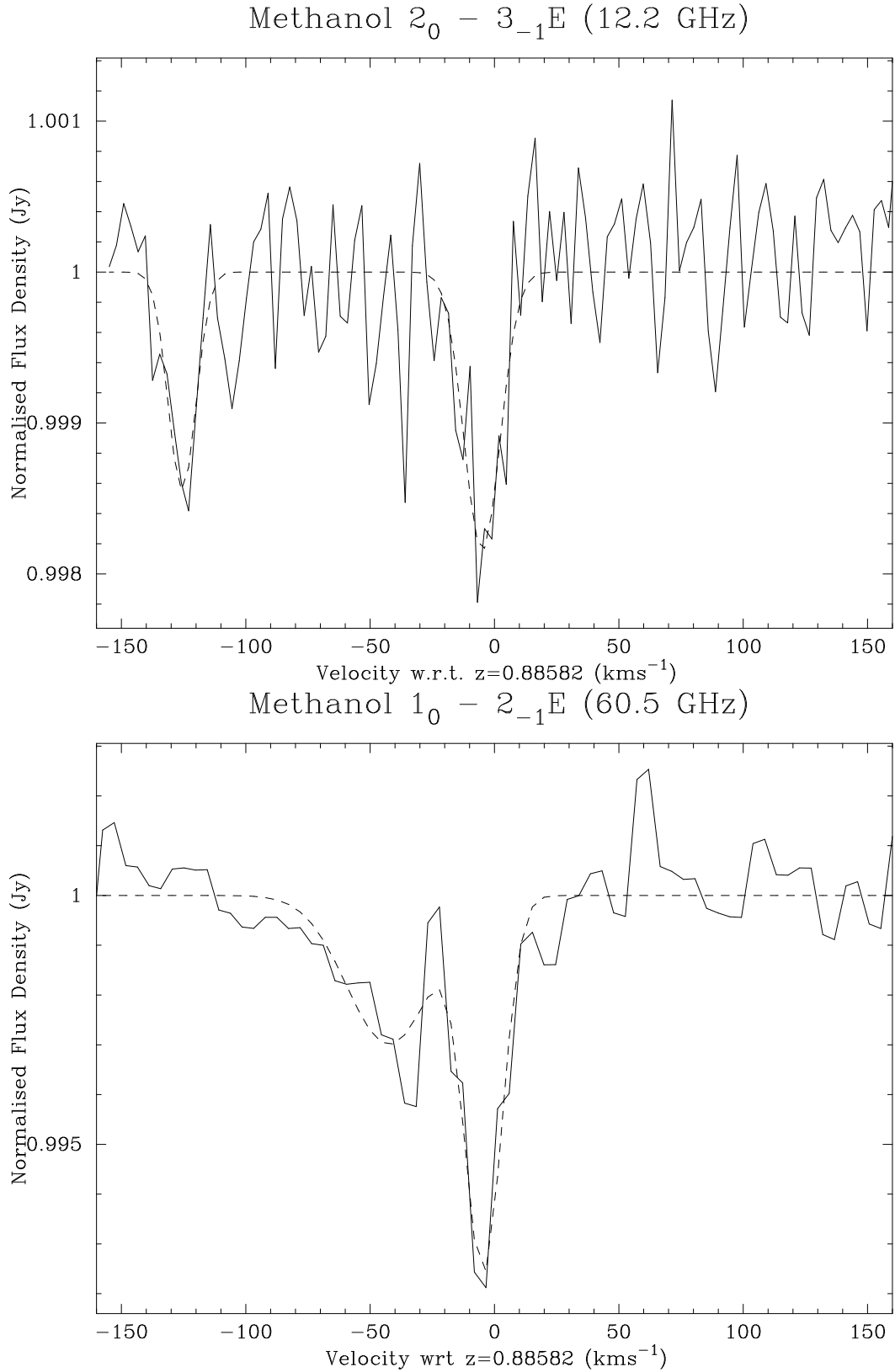

\begin{center}
      \includegraphics[angle=270,scale=0.65]{meth12.ps}
      \includegraphics[angle=270,scale=0.65]{meth60.ps}
\end{center}
\caption{{\em Top:} The absorption from the 12.2~GHz $2_{0} \rightarrow 3_{-1}E$ transition of methanol towards PKS\,B1830$-$211.  {\em Bottom:} The absorption from the 60.5~GHz $1_{0} \rightarrow 2_{-1}E$ transition over the same velocity range as the 12.2 GHz transition \citep{Muller+11}.}  \label{fig:abs}
\end{figure}

The normalised absorption spectrum for the 12.2~GHz transition of methanol towards the south-western component of PKS\,B1830$-$211 is shown in Figure~\ref{fig:abs}.  The maximum absorption is only about 5 times the RMS noise in the average spectrum, but is spectrally well resolved, and was detected in each of the individual observing sessions.  The maximum absorption is very close to the velocity at which the strongest absorption is seen in previous molecular absorption studies towards PKS\,B1830$-$211 \citep[e.g.][]{Wiklind+96,Muller+06,Henkel+08,Muller+11}, with additional, slightly weaker absorption observed at approximately -130~\kms.  Previous interferometric studies have shown that the absorption close to 0~\kms\/ arises from the south-western component of the gravitational lense, while that at around -150~\kms\/ arises from the north-eastern component \citep{Muller+06}.  

No absorption was detected from the 19.9 GHz transition, with a 3-$\sigma$ limit on the level of absorption in the normalised flux density of 0.019.  The large RMS compared to the observation of the 12.2 GHz transition is because 10.5 GHz lies well beyond the nominal bounds of the current ATCA X-band system and the system performance is very poor compared to that below 9~GHz.

\section{Discussion}

The primary purpose of these observations was to use the sensitivity of different rotational transitions of the methanol molecule to investigate if there is any evidence that the proton-to-electron mass ratio was different at a redshift of 0.89 to that observed at the current epoch.  We have fitted a single Gaussian profile to each of the two 12.2 GHz methanol absorption features.  The spectral profiles of each of these may be more complex, however, with the signal to noise ratio of the current observations we are only justified in fitting a single component.  The parameters of the fitted Gaussians are given in Table~\ref{tab:gauss}.  We have also fitted Gaussian profiles to the spectrum of the 60.5 GHz methanol transition observed by \citet{Muller+11}\footnotemark.  The 60.5 GHz transition (see Fig~\ref{fig:abs}) also shows its strongest absorption near $-5$~\kms, however, there is a secondary component with about half the optical depth at $-40$~\kms.  There is a hint of this component in the 12.2 GHz spectrum, but significantly better signal to noise would be required to reliably establish if this component is present.  The HCO$^+$ and HCN spectra show weaker absorption covering this velocity range \citep{Muller+08,Muller+11}.  Unlike the 12.2 GHz spectrum, there is no component in the 60.5 GHz spectrum associated with the north-eastern component of the gravitational lens.  This likely reflects differences in the morphology of the background continuum source at the two frequencies, as the expectation is that the source size is greater at lower frequencies (due to the steep spectrum of the jet).  Examination of the 60.5 GHz spectrum shows that the channels appear paired, with every second channel very similar to its predecessor.  The observations of \citet{Muller+11} had a spectral resolution of 1~MHz, but the spectra are displayed with a channel separation of 0.5~MHz and this appears to be the underlying reason behind the close correlation between consecutive channels (this can also be seen in many of the spectra displayed in \citet{Muller+11}, e.g. their figure~7).  The effective velocity resolution (i.e. spectral resolution of 1 MHz) of the 60.5 GHz spectrum in the rest frame of the absorbing system is 9.4~\kms.

\footnotetext[1]{The 60.5 GHz spectrum was extracted from the ascii format spectrum obtained from http://cdsarc.u-strasbg.fr/viz-bin/qcat?J/A+A/535/A103}

We have calculated the optical depth of the absorption for each transition (see Table~\ref{tab:gauss}) using equation 1 of \citet{Muller+11}, assuming 38\% of the total flux density from the source originates from the south-western component.  We are not able to measure this fraction through our observations, but rather use the value derived from earlier observations of saturated lines by \citet{Muller+06,Muller+11}. This value is also consistent with imaging studies which directly measure the flux density in the two components \citep[e.g.][]{Nair+93,Lovell+98}. Temporal and spectral variations in the source mean that this factor is somewhat uncertain for individual epochs, but it is the best estimate available.  The difference in the central velocity of the main absorption components from the two transitions is $-0.6 \pm 1.6$ \kms, with the uncertainty calculated by adding the formal error from the two Gaussian fits, and other sources of error in the relative velocity scales in quadrature.  The uncertainties in the rest frequency of the two methanol transitions and in the Barycentric velocity corrections for both the current observations and the \citeauthor{Muller+11} spectrum, added in quadrature contribute approximately $0.13$~\kms\/ to the total error budget.

The sensitivity coefficients for calculating changes in the proton-to-electron mass ratio for the 12.2 and 60.5 GHz methanol transitions are $K_\mu$ = -33 and -7.4 respectively \citep{Jansen+11,Levshakov+11}.  Using the observed difference between the velocity of the absorption in these two transitions we can constrain the proton-to-electron mass ratio $\Delta \mu/\mu = 0.8 \pm 2.1 \times 10^{-7}$, corresponding to a 3-$\sigma$ limit of $\Delta \mu/\mu < 6.3 \times 10^{-7}$.  This is a factor of 2 better than previous constraints obtained from ammonia observations in the same source \citep{Henkel+09}, and comparable to the 3-$\sigma$ limit of $3.6 \times 10^{-7}$ derived from simultaneous fitting of multiple molecular transitions in B0218+357 at $z$ = 0.685 \citep{Kanekar11}.  We note, that our observations are not consistent with 4-$\sigma$ detection of a variation in $\mu$ by \cite{Muller+11} of $\Delta \mu/\mu = -1.95 \pm 0.47 \times 10^{-6}$ obtained by comparing a large number of different molecular transitions with absorption near 0~\kms.  The look-back time for the $z$=0.89 absorbing system is 7.24 Gyr (using standard $\Lambda$CDM with $H_0$ = 71\kms Mpc$^{-1}$ ; $\Omega_m = 0.27$ ; $\Omega_\Lambda = 0.73$), this translates to an upper limit on the time variation in the proton-to-electron mass ratio of $|\dot{\mu}/\mu| = 8.7 \times 10^{-17}$ yr$^{-1}$.

The two transitions observed here are from the same family, so we would expect the intrinsic absorption profiles of the two lines to be similar, with differences likely reflecting a differences in the morphology of the background continuum source at the two frequencies.  The absorption in the 12.2 GHz~methanol transition is approximately a factor of 4 lower in optical depth than that seen in the 60.5 GHz transition.  This is likely because the energy of the lower-state for the 60.5 GHz transition is 12.5~K above the ground state, compared to 19.5 K for the 12.2 GHz transition.  This is consistent with earlier observations of molecular absorption in PKS\,B1830$-$211 which find that the detections are primarily from transitions to the ground state, or those with low energy levels \citep{Muller+11}.  Observations covering the frequency of three other methanol transitions are also available, the 19.9 GHz ($2_{1} \rightarrow 3_{0}E$) transition was observed as part of the current project, but not detected, with a 3$\sigma$ limit on the optical depth of $<$0.05, and the 68.3 GHz ($1_{1} \rightarrow 2_{0}E$) and 84.5 GHz ($5_{-1} \rightarrow 4_{0}E$) transitions were within frequency coverage of the spectral scan of \citet{Muller+11}, for which the 3$\sigma$ limits on the optical depth are $<$ 0.004 and $<$0.006 respectively.  These three transitions have lower-states 27, 20 and 36 K above the ground-state for the 19.9, 68.3 and the 84.5 GHz respectively, but also have weaker line strengths than either the 12.2 or 60.5 GHz transitions, so their non-detection is perhaps not surprising.  The FWHM of the main component of absorption is 17~\kms\/ for both the 12.2 and 60.5 GHz transitions.  This is significantly larger than the typical values observed in 12.2 GHz absorption seen towards individual cold clouds in the Galaxy, which have a median of 4.2~\kms\/ \citep{Peng+92}.  Within the Milky Way, only clouds near the Galactic Centre have widths that approach or exceed 17~\kms, which suggests that the main absorption component may be a blend of two or more clouds along the line of sight from the south-western component.

Formally, the strongest constraints on variations in the proton-to-electron mass ratio which have been reported to date are those from \citet{Kanekar11}, obtained through simultaneous multi-component Voigt fitting of three blended absorption systems from high signal-to-noise ratio spectra from three different molecular species (NH$_{3}$, CS and H$_2$CO), which yielded  $|\Delta \mu/\mu| = -3.5 \pm 1.2 \times 10^{-7}$.  That the measurement is approximately three times the quoted uncertainty may be statistical chance, or it may indicate that the quoted uncertainty is an underestimate (for example the degree to which the absorption from different molecules is co-spatial is very difficult to accurately quantify).  

\section*{Conclusions}

We have undertaken observations of the 12.2~GHz transition of methanol towards the $z$=0.89 absorption system in PKS\,B1830$-$211.  Combining our results with the only other published methanol absorption spectrum in this source we have been able to constrain changes in the proton-to-electron mass ratio over the last 7.24 Gyr to be $< 6.3 \times 10^{-7}$ (3-$\sigma$).  We have achieved a similar accuracy to the most precise previous cosmological measurements of $\Delta \mu/\mu$, but through much more direct and simpler means (fitting only a single component to two transitions of the same molecule).  Future higher sensitivity observations of the methanol absorption are likely to reveal additional components, however, within Galactic clouds methanol is less widely distributed than molecules such as HCO$^+$ or HCN.  Hence, the relative simplicity of the methanol absorption spectra in PKS\,B1830$-$211 compared to that observed in other molecules, combined with the unique sensitivity of rotational transitions of methanol to changes in the proton-to-electron mass ratio, identified by \citet{Jansen+11} and \citet{Levshakov+11}, suggest that future observations with better signal to noise and/or including additional transitions should be able to either measure changes in $\mu$, or constrain it to levels better than 1 part in 10 million (i.e. a factor of 5 or more better than the observations presented here).  In particular the $1_{0} \rightarrow 0_{0}A^{+}$ transition, which has a rest frequency of 48.4~GHz should be readily detectable (redshifted to 25.7 GHz) using either the ATCA or EVLA, although possible blending with the $1_{0} \rightarrow 0_{0}E$ transition (separated by 4.4 MHz or $\sim$28~\kms), may cause additional complexity.

\section*{Acknowledgements}

We thank an anonymous referee for comments which have improved this paper.  The Australia Telescope Compact Array is part of the Australia Telescope which is funded by the Commonwealth of Australia for operation as a National Facility managed by CSIRO.  This research has made use of NASA's Astrophysics Data 
System Abstract Service.

\clearpage

\begin{table}
\begin{center}
\caption{Summary of observations.  Column 5 gives the time on-source for PKS\,B1830$-$211.  Column 7 gives the Barycentric velocity of the observatory towards PKS\,B1830-211 at the time of the start and end of the observation.} \label{tab:obs}
\begin{tabular}{ccccccc}
\tableline
\multicolumn{1}{c}{\bf Date} & \multicolumn{1}{c}{\bf Time} & \multicolumn{1}{c}{\bf Array} & \multicolumn{1}{c}{\bf Correlator} & \multicolumn{1}{c}{\bf Time on-} & \multicolumn{1}{c}{\bf Continuum} & \multicolumn{1}{c}{\bf Barycentric}\\
\multicolumn{1}{c}{\bf (2011)} & \multicolumn{1}{c}{\bf (UT)} & \multicolumn{1}{c}{\bf Config.} & \multicolumn{1}{c}{\bf Config.} & \multicolumn{1}{c}{\bf source (min)} & \multicolumn{1}{c}{\bf Flux (Jy)} &\multicolumn{1}{c}{\bf Velocity (\kms)} \\
\tableline
November 4 & 08:00 -- 11:15 & 750C & CFB 1M-0.5k & 135 & 9.640 & 24.98,25.10 \\
November 18 & 03:00 -- 04:30 & 1.5D & CFB 64M-32k & 49 & 7.785 & 19.96,20.04 \\
December 7 & 05:40 -- 06:40 & 6.0A & CFB 64M-32k & 60 & 9.740 & 11.87,11.95 \\
\tableline
\end{tabular}
\end{center}
\end{table}

\clearpage

\begin{table}
\begin{center}
\caption{Gaussian fits to absorption spectra} \label{tab:gauss}
\begin{tabular}{ccccccc}
\tableline
\multicolumn{1}{c}{\bf Transition} & \multicolumn{1}{c}{\bf Velocity} & \multicolumn{1}{c}{\bf $\tau$} & \multicolumn{1}{c}{\bf FWHM} \\
                                                            & \multicolumn{1}{c}{\bf (\kms)} &  & \multicolumn{1}{c}{\bf (\kms)} \\
\tableline
12.2 GHz & -5.0 $\pm$ 1.3 & 0.0047 & 17.0 $\pm$ 2.9 \\
                  & -125.1 $\pm$ 2.1 & 0.0037 & 13.2 $\pm$ 3.3 \\
60.5 GHz & -4.4 $\pm$ 0.9 & 0.0196 & 17.2 $\pm$ 2.0 \\
                  & -42.4 $\pm$ 3.4 & 0.0079 & 40 $\pm$ 9 \\
\tableline
\end{tabular}
\end{center}
\end{table}

\clearpage

\begin{thebibliography}{}


\bibitem[Breen et al.(2011)]{Breen+11}
  Breen, S.L., Ellingsen, S.P., Caswell, J.L., Green, J.A., Fuller, G.A., Voronkov, M.A. Quinn L.J. \& Avison, A., 2011, MNRAS, in press
  
\bibitem[Carilli et al.(1998)]{Carilli+98}
  Carilli, C.L., Menten, K.M., Reid, M.J., Rupert, M.P. \& Claussen, M., 1998, in Petitjean, P., Charlot, S., eds, 13th Colloque d'Astrophysique de l'Institut d'Astrophysique de Paris, Structure et Evolution du Milieu Inter-Galactique Revele par Raies D'Absorption dans le Spectre des Quasars, p. 325
  
\bibitem[Chengalur et al.(1999)]{Chengalur+99}
  Chengalur, J.N., de Bruyn, A.G. \& Narasimha, D., 1999, A\&A, 343, L79

\bibitem[Ellingsen et al.(2011)]{Ellingsen+11}
  Ellingsen, S.P., Voronkov, M.A \& Breen, S.L., 2011, Phys. Rev. Lett., 107, 270801  
    
\bibitem[Flambaum \& Kozlov(2007)]{Flambaum+07}
  Flambaum, V.V. \& Kozlov, M.G., 2007, Phys. Rev. Lett., 98, 240801

\bibitem[Henkel et al.(2008)]{Henkel+08}
  Henkel, C., Braatz, J.A., Menten, K.M. \& Ott, J., 2008, A\&A, 485, 451

\bibitem[Henkel et al.(2009)]{Henkel+09}
  Henkel, C., Menten, K.M., Murphy, M.T., Jethava, N., Flambaum, V.V., Braatz, J.A., Muller, S., Ott, J. \& Mao, R.Q., 2009, A\&A, 500, 725

\bibitem[Jansen et al.(2011)]{Jansen+11}
 Jansen, P., Xu, L.-H., Kleiner, I., Ubachs, W. \& Bethlem, H.L., 2011, Phys. Rev. Lett., 106, 100801

\bibitem[Jauncey et al.(1991)]{Jauncey+91}
  Jauncey, D.L. et al., 1991, Nat, 352, 132
  
\bibitem[Jones et al.(1996)]{Jones+96}
  Jones, D.L. et al., 1996, ApJ, 470, L23

\bibitem[Kanekar(2011)]{Kanekar11}
 Kanekar, N., 2011, ApJ, 728, L12

\bibitem[Kanekar et al.(2005)]{Kanekar+05}
 Kanekar, N., et al. 2005, Phys. Rev. Lett., 95, 1301
 
\bibitem[Kanekar et al.(2010)]{Kanekar+10}
 Kanekar, N., Chengalur, J.N. \& Ghosh, T., 2010, ApJ, 716, L23

\bibitem[King et al.(2008)]{King+08}
 King, J.A., Webb, J.K., Murphy, M.T. \& Carswell, R.F., 2008, Phys. Rev. Lett., 101, 251204
 
\bibitem[Levshakov et al.(2006)]{Levshakov+06}
 Levshakov, S.A.,  Centuri\'on, M., Molaro, P., D'Odorico, S., Reimers, D., Quast, R. \& Pollmann, M., 2006, A\&A, 449, 879
 
\bibitem[Levshakov et al.(2011)]{Levshakov+11}
 Levshakov, S.A., Kozlov, M.G. \& Reimers D., 2011, Astrophys. J., 738, 26
 
 \bibitem[Lidman et al.(1999)]{Lidman+99}
   Lidman, C. et al., 1999, ApJ, 514, 57
 
 \bibitem[Lovell et al.(1996)]{Lovell+96}
   Lovell, J.E.J. et al., 1996, ApJ, 472, 5

 \bibitem[Lovell et al.(1998)]{Lovell+98}
   Lovell, J.E.J. et al., 1998, ApJ, 508, L51
      
 \bibitem[Menten(1991)]{Menten91}
   Menten, K.M., 1991, ApJ, 380, L75

\bibitem[M\"uller et al.(2004)]{Muller+04}
  M\"uller, H.S.P, Menten, K.M. \& M\"ader, H., 2004, A\&A, 428, 1019
  
\bibitem[Muller et al.(2006)]{Muller+06}
  Muller, S., Gu\'elin, M., Dumke, M.,Lucas, R. \& Combes, F., 2006, A\&A, 458, 417

\bibitem[Muller \& Gu\'elin(2008)]{Muller+08}
  Muller, S. \& Gu\'elin, M., 2008, A\&A, 491, 739

\bibitem[Muller et al.(2011)]{Muller+11}
  Muller, S., Beelen, A., Gu\'elin, M., Aalot, S., Black, J.H., Combes, F., Curran, S.J., Theule, P. \& Longmore, S.N., 2011, A\&A, 535, 103
  
\bibitem[Nair et al.(1993)]{Nair+93}
  Nair, S., Narasimha, D. \& Rao, A.P., 1993, ApJ, 407, 46
  
\bibitem[Peng \& Whiteoak(1992)]{Peng+92}
  Peng, R.S. \& Whiteoak, J.B., 1992, 254, 301

\bibitem[Sault et al.(1995)]{Sault+95}
  Sault, R.J., Teuben, P.J. \& Wright, M.H., 1995, in Shaw R.A., Payne H.E., Hayes, J.J.E., eds, ASP Conf. Ser. Vel. 77, Astronomical Data Analysis Software and Systems IV. Astron. Soc. Pac., San Francisco, p. 433
 
\bibitem[Subrahmanyan et al.(1990)]{Subrahmanyan+90}
  Subrahmanyan, R., Narasimha, D., Pramesh Rao, A. \& Swarup, G., 1990, MNRAS, 246, 263
  
\bibitem[van der Tak et al.(2000)]{vanderTak+00}
  van der Tak, F.F.S, van Dishoeck, E.F., Caselli, P., 2000, A\&A, 361, 327
  
\bibitem[Webb et al.(1999)]{Webb+99}
  Webb, J.K., Flambaum, V.V., Churchill, C.W., Drinkwater, M.J. \& Barrow, J.D., 1999, Phys. Rev. Lett., 82, 884
  
\bibitem[Wilson et al.(2011)]{Wilson+11}
  Wilson, W.E. et al., 2011, MNRAS, 416, 832

\bibitem[Wiklind \& Combes(1996)]{Wiklind+96}
  Wiklind, T. \& Combes, F., 1996, Nat, 379, 139
 
 \bibitem[Winn et al.(2002)]{Winn+02}
  Winn, J.N. et al., 2002, ApJ, 575, 103
   
  \bibitem[Uzan(2011)]{Uzan11}
    Uzan, J.-P., 2011, Living Rev. Relativity, 14, 2

\end{thebibliography}
\end{document}